\documentclass[aip,jcp,reprint]{revtex4-1}
\usepackage{amsmath}
\usepackage{graphicx}
\usepackage{tikz,multirow,tabularx}
\usepackage{footmisc}
\newcommand{\vect}[1]{\vec{\mathrm{#1}}}

\newcommand{\mol}[1]{\mathrm{#1}}
\newcommand{\ibo}{\phi}

\newcommand{\weight}{w}
\newcommand{\Nocc}{N_{\mathrm{occ}}}
\newcommand{\Natom}{N_{\mathrm{atom}}}

\newcommand{\mat}[1]{\mathbf{#1}}
\newcommand{\vecrep}{\vec{\mathrm{v}}}

\newcommand{\Dsq}{D^{2}}
\newcommand{\Klin}{K_{\mathrm{lin}}}
\newcommand{\KGauss}{K_{\mathrm{Gauss}}}
\newcommand{\Kmol}{K}
\newcommand{\sigmaresc}{\sigma_{\mathrm{resc}}}

\newcommand{\dimf}{D_{f}}
\newcommand{\sigmaglobal}{\sigma_{\mathrm{global}}}
\newcommand{\dpair}[3]{\mol{#1}_{#2\rightarrow#3}}
\newcommand{\rcut}{\rho_{\mathrm{cut}}}

\newcommand{\Ntrain}{N_{\mathrm{train}}}
\newcommand{\lmax}{l_{\mathrm{max}}}
\usepackage{dcolumn}
\newcommand{\alignedmultrow}[3]{\multirow{#1}{*}{
\begin{tabular}{D{,}{}{#2}}
    #3
\end{tabular}}}
\begin{document}
\date{\today}
%\title{An orbital-based representation of changes in electronic structure for improved $\Delta$-machine learning approaches}
\title{An orbital-based representation for accurate Quantum Machine Learning}

\author{Konstantin Karandashev}
\email{konstantin.karandashev@univie.ac.at}
\affiliation{University of Vienna, Faculty of Physics, Kolingasse 14-16, AT-1090 Wien, Austria}
\author{O. Anatole von Lilienfeld}
\email{anatole.vonlilienfeld@univie.ac.at}
\affiliation{University of Vienna, Faculty of Physics, Kolingasse 14-16, AT-1090 Wien, Austria}
\affiliation{Institute of Physical Chemistry and National Center for Computational Design and Discovery of Novel Materials (MARVEL), Department of Chemistry, University of Basel, Klingelbergstrasse 80, CH-4056 Basel, Switzerland}

\begin{abstract}

We introduce an electronic structure based representation for quantum machine learning (QML) of electronic properties throughout chemical compound space. The representation is constructed using computationally inexpensive {\it ab initio} calculations and explicitly accounts for changes in the electronic structure. We demonstrate the accuracy and flexibility of resulting QML models  when applied to property labels such as total potential energy, HOMO and LUMO energies, ionization potential, and electron affinity, using as data sets for training and testing entries from the QM7b, QM7b-T, QM9, and LIBE libraries. For the latter, we also demonstrate the ability of this approach to account for molecular species of different charge and spin multiplicity, resulting in QML models that infer total potential energies based on geometry, charge, and spin as input.
    
\end{abstract}

\maketitle

\section{Introduction}
\label{sec:introduction}

Increasing the electrochemical stability windows of electrolytes used in energy storage devices is critical for improving the energy density, as it grows with operating voltage (linearly for batteries and quadratically for electrical double layer capacitors).\cite{Borodin:2019} One aspect of this problem consists of searching chemical compound space for optimal organic solvents, often addressed with high-throughput compute campaign protocols based on {\it ab initio} calculations and common physically motivated approximations.\cite{Korth:2014,Cheng_Curtiss:2015,Borodin_Knap:2015,Qu_Persson:2015,Lian_Wu:2019} The computational cost associated with such screening efforts can be drastically decreased using modern machine learning techniques.\cite{Wang_Gomez-Bombarelli:2020} It has also been shown\cite{Zubatyuk_Isayev:2021} that machine learning can yield impressive accuracy on predicting energy changes associated with attaching and detaching an electron from a molecule, which corresponds to a first order  approximation\cite{Borodin:2019} of the electrochemical stability windows.  Alas, satisfying accuracy can only be achieved for very large training datasets, implying the need for machine learning models with improved learning efficiency.

A compromise between estimating a quantity with accurate but costly {\it ab initio} calculations and fast but data hungry conventional machine learning models could be realized if inexpensive {\it ab initio} calculations can be exploited to improve the performance of machine learning model. Examples of this philosophy include using transfer learning techniques on a neural network trained with abundant but inaccurate data to re-train it with accurate but scarce data\cite{Grambow_Green:2019,Smith_Roitberg:2019} and using machine learning to automatically tune parameters of a semiempirical calculation.\cite{Dral_Thiel:2015} A more straightforward approach is to use $\Delta$-machine learning\cite{Ramakrishnan_Lilienfeld:2015,Zaspel_Lilienfeld:2019,Mezei_Lilienfeld:2020} ($\Delta$-ML), {\it i.e.} use machine learning to predict the difference between a quantity and its estimate from a relatively inexpensive calculation; this is the simplest example of a multifidelity information fusion approach.\cite{Batra_Ramprasad:2019} $\Delta$-ML approaches can be further improved by incorporating additional features from the base-line calculations ({\it e.g.} orbital coefficients, operator matrices) into a model's input; such frameworks have performed well in conjunction with Hartree-Fock,\cite{Welborn_Miller:2018,Cheng_Miller:2019a,Cheng_Miller:2019b,Chen_E:2020} DFT,\cite{Dick_Fernandez-Serra:2019,Dick_Fernandez-Serra:2020} and semiempirical\cite{Qiao_Miller:2020,Christensen_Miller:2021,Qiao_Miller:2021} methods.

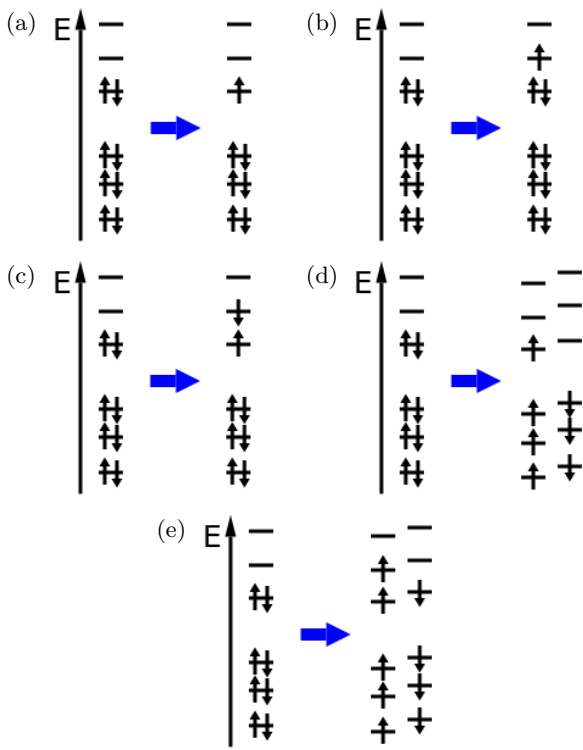
\begin{figure}[t]
\begin{tabularx}{\textwidth}{cc}
    \begin{tikzpicture}
            \node[anchor=south west,inner sep=0] (image) at (0,0) {\includegraphics[width=0.18\textwidth]{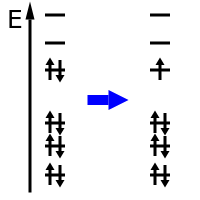}};
            \node at (-0.3,3.0) {(a)};
        \end{tikzpicture}
&
    \begin{tikzpicture}
            \node[anchor=south west,inner sep=0] (image) at (0,0) {\includegraphics[width=0.18\textwidth]{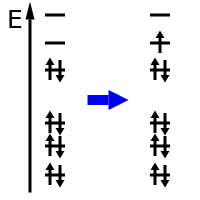}};
            \node at (-0.3,3.0) {(b)};
        \end{tikzpicture}
\\
    \begin{tikzpicture}
            \node[anchor=south west,inner sep=0] (image) at (0,0) {\includegraphics[width=0.18\textwidth]{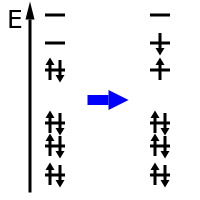}};
            \node at (-0.3,3.0) {(c)};
        \end{tikzpicture}
&
    \begin{tikzpicture}
            \node[anchor=south west,inner sep=0] (image) at (0,0) {\includegraphics[width=0.18\textwidth]{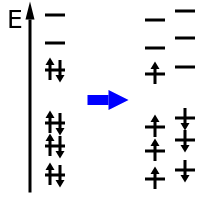}};
            \node at (-0.3,3.0) {(d)};
        \end{tikzpicture}
\\
\multicolumn{2}{c}{\begin{tikzpicture}
            \node[anchor=south west,inner sep=0] (image) at (0,0) {\includegraphics[width=0.18\textwidth]{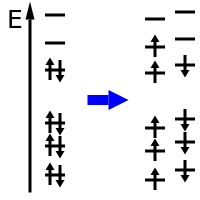}};
            \node at (-0.3,3.0) {(e)};
        \end{tikzpicture}}
        \\
\end{tabularx}
  \caption{\label{fig:Slater_determinant_change} Changes of Slater determinants used in this work to generate representations used for modelling (a) HOMO energy, (b) LUMO energy, (c) first excitation, (d) ionization potential, and (e) electron affinity.}
\end{figure}

In this work, we introduce an extension of the latter approach. We use Hartree-Fock calculations to either map a system's wavefunction or change of a system's wavefunction on a Slater determinant or pairs of Slater determinants. 
Since the underlying representations are generated from localized orbital coefficients and matrices $\mat{F}$ (Fock), $\mat{J}$ (Coulomb), and $\mat{K}$ (exchange), we dub the proposed representation \emph{FJK}. Mapping onto single Slater determinants is used to estimate a system's energy, and in this work we used it to predict both ground state and spin-excited state energies. To estimate energy changes associated with changes of electronic structures we use mapping onto pairs of Slater determinants that are physically more representative of such changes. The wavefunction pairs used in this work are illustrated in Figure~\ref{fig:Slater_determinant_change}, with a more detailed explanation given in Subsec.~\ref{subsec:orbital_based_kernel}.

Apart from introducing the representation, we have also tested FJK on various properties and systems, including ionization potentials, electron affinities, as well as HOMO and LUMO energies available for molecular entries in the QM7b\cite{Blum_Reymond:2009,Montavon_Lilienfeld:2013} and QM9\cite{Ruddigkeit_Reymond:2012,Ramakrishnan_Lilienfeld:2014} datasets. 
All these four quantities are assumed to be related to electrochemical stability windows to a reasonable extent.\cite{Borodin:2019} We also tested FJK's performance for predicting total potential energies of molecules in QM7b-T,\cite{Welborn_Miller:2018,Cheng_Miller:2019a} GDB-13-T,\cite{Welborn_Miller:2018,Cheng_Miller:2019a} and LIBE\cite{Spotte-Smith_Persson:2021} datasets. QM7b-T and GDB-13-T have been extensively used as benchmarks for testing quantum machine learning models' ability to predict total potential energies, while LIBE dataset is of particular interest for two reasons. Firstly, it was generated to be representative of the kind of species that could play a role in solid electrolyte interphase formation in lithium batteries, a process that affects greatly a battery's performance and longevity.\cite{Winter:2009,Verma_Novak:2010,An_Wood_III:2016} Secondly, it contains species of different charge and spin states, enabling us to test FJK's ability to process them in a single model, which is an interesting possibility as also recently reported for AIMNet-NSE.\cite{Zubatyuk_Isayev:2021}

The paper is structured as follows. In Sec.~\ref{sec:theory} we describe the theory behind FJK. In Secs.~\ref{sec:global_comp_details} and~\ref{sec:datasets} we provide technical details of numerical experiments and datasets used for this work, and with numerical results presented in Sec.~\ref{sec:results_discussion}. In Sec.~\ref{sec:computational_timing}, we discuss the computational cost associated with the proposed method and compare to conventional $\Delta$-ML approaches. Section~\ref{sec:conclusions_outlook} summarizes the paper and outlines possible directions of future work.

\section{Theory}
\label{sec:theory}

\subsection{General idea}
\label{subsec:general_idea}

We rely on kernel ridge regression,\cite{Vapnik:1998} a supervised machine learning approach that, when applied to chemical problems, can be used to infer  $p_{\mathrm{est}}$ as an estimate of property $p$ for any given query compound with representation $q$ as
\begin{equation}
    p_{\mathrm{est}}(q)=\sum_{t=1}^{\Ntrain}\vec{\alpha}_{t}K(q, q_{t}),
\label{eq:krr_prediction}
\end{equation}
where $\Ntrain$ is the size of the training set, $q_{t}$ ($t=1,\ldots,\Ntrain$) are training set compounds, $K$ is the kernel function which quantifies the similarity between query and test compound. 
The vector $\vec{\alpha}$ is calculated as
\begin{equation}
    \vec{\alpha}=(\mat{K}+\lambda\mat{I})^{-1}\vect{P},
    \label{eq:alpha}
\end{equation}
where $\mat{I}$ is the identity matrix, $\lambda$ is a hyperparameter reflecting the level of noise (numerical or statistical) associated with the data, $\mat{K}$ and $\vect{P}$ are the kernel matrix and training quantity vector defined as
\begin{align}
    \mat{K}_{t't''}=&K(q_{t'},q_{t''}),\\
    \vect{P}_{t'}=&p(q_{t'}),
\end{align}
with $t'$ and $t''$ taking values from $1$ to $\Ntrain$.

The error of estimating $p(q)$ can be decreased using a physically motivated approximate estimate $p^{\mathrm{approx}}$ that is inexpensive to evaluate. The idea of $\Delta$-ML methods\cite{Ramakrishnan_Lilienfeld:2015} is to choose $p^{\mathrm{approx}}(q)$ such that the error of estimating $p(q)-p^{\mathrm{approx}}(q)$ is smaller than the error of estimating $p(q)$ for a given $\Ntrain$; this approach has more sophisticated generalizations for cases when several approximations of differing cost and accuracy are available.\cite{Zaspel_Lilienfeld:2019} A natural extension of the concept is to use byproducts of calculating $p^{\mathrm{approx}}$ to define a representation of compound $q$ that would reflect not just the compound's features, but also physical intuition behind the property $p$. The general idea of the method proposed in this work is to define representations for localized orbitals obtained from a Hartree-Fock calculation and then define the kernel function in terms of these orbital representations obtained from the ground state or excited state calculations. This concept has similarities to the one of MOB-ML,\cite{Welborn_Miller:2018,Cheng_Miller:2019a,Cheng_Miller:2019b} and we will do a more thorough comparison of the two approaches in the end of this Section.

\subsection{FJK representation}
\label{subsec:orbital_based_representation}

We will now explain how we use FJK to represent localized orbitals of a Slater determinant $\mol{A}$ which we will refer to as $\ibo_{i}^{\mol{A}}$ ($i=1,\ldots,\Nocc^{\mol{A}}$, where $\Nocc^{\mol{A}}$ is the number of occupied orbitals). In this work, we used intrinsic bond orbitals\cite{Knizia:2013} as they provide a straightforward path to analyzing a Slater determinant in terms of chemically intuitive concepts, which have been shown to be very useful for HOMO and LUMO energy predictions.\cite{Mazouin_Lilienfeld:2021}

We denote a coefficient $\ibo_{i}^{\mol{A}}$ with respect to an atomic orbital as $C_{kln}^{\mol{A},i}$, where $k$ is the index of the atom on which the atomic orbital is centered ($k=1,\ldots,\Natom^{\mol{A}}$, where $\Natom^{\mol{A}}$ is the number of atoms), $l$ is the angular momentum number corresponding to the atomic orbital ($l=1,\ldots,\lmax$, where $\lmax$ is the maximum angular momentum number represented in the basis), $n$ an index denoting a distinct combination of all other parameters of an atomic orbital. We will assign each contribution to $\ibo_{i}^{\mol{A}}$ from atomic orbitals centered on a given atom $k$ a nonnegative weight and a vector representation. The weight function is defined as
\begin{equation}
 \weight_{k}(\ibo_{i}^{\mol{A}})=\sum_{l=0}^{\lmax}\sum_{n1,n2}S_{kln_{1},kln_{2}}C_{kln_{1}}^{\mol{A},i}C_{kln_{2}}^{\mol{A},i},
\end{equation}
where $S$ is the overlap matrix. Due to the localization procedure some $\weight_{k}(\ibo_{i}^{\mol{A}})$ are negligibly small; in order to smoothly cut off such weights we will also use $\tilde{\weight}_{k}(\ibo_{i}^{\mol{A}})$ defined as
\begin{equation}
    \tilde{\weight}_{k}(\ibo_{i}^{\mol{A}}):=\mathrm{max}[\weight_{k}(\ibo_{i}^{\mol{A}})-\rho_{-}, 0],
    \label{eq:cut_weights_def}
\end{equation}
with $\rho_{-}$ such that
\begin{equation}
    \frac{\sum_{k=1}^{\Natom^{\mol{A}}}\tilde{\weight}_{k}(\ibo_{i}^{\mol{A}})}{\sum_{k=1}^{\Natom^{\mol{A}}}\weight_{k}(\ibo_{i}^{\mol{A}})}=1-\rcut,
\end{equation}
where $\rcut$ is a parameter chosen by the user.

The vector representation of a contribution to $\ibo_{i}^{\mol{A}}$ from atomic orbitals centered on atom $k$ is defined in terms of the following ``coupling functions''
\begin{align}
s^{k,M}_{l_{1}l_{2}}(\ibo_{i}^{\mol{A}})=&\frac{1}{\weight_{k}(\ibo_{i}^{\mol{A}})}\sum_{n_{1},n_{2}} M_{kl_{1}n_{1},kl_{2}n_{2}}C_{kl_{1}n_{1}}C_{kl_{2}n_{2}},\label{eq:self_coupling}\\
o^{k,M}_{l_{1}l_{2}}(\ibo_{i}^{\mol{A}})=&\frac{1}{\weight_{k}(\ibo_{i}^{\mol{A}})}\sum_{\substack{m=1 \\ m\neq k}}^{\Natom^{\mol{A}}}\sum_{n_{1},n_{2}}M_{kl_{1}n_{1},ml_{2}n_{2}}C_{kl_{1}n_{1}}C_{ml_{2}n_{2}},\label{eq:other_coupling}
\end{align}
where  $M$ is a matrix defined over atomic orbitals. The final vector representation $\vecrep^{k}(\ibo_{i}^{\mol{A}})$, whose dimensionality we will denote as $\dimf$, consists of the following components
\begin{equation}
\begin{split}
    \vecrep^{k}(\ibo_{i}^{\mol{A}})=&\{s^{k,S}_{ll}(\ibo_{i}^{\mol{A}}): l=1,\ldots,\lmax;\\
&    s^{k, M}_{l_{1}l_{2}}(\ibo_{i}^{\mol{A}}): l_{1}=0,\ldots,\lmax,\\
& l_{2}=0,\ldots,\lmax, l_{1}\leq l_{2}, M=F,J,K;\\
&o^{k, M}_{l_{1}l_{2}}(\ibo_{ik}^{\mol{A}}): l_{1}=0,\ldots,\lmax,\\
&l_{2}=0,\ldots,\lmax, M=F, J, K\},
\end{split}
\end{equation}
where $F$, $J$, and $K$ are Fock, Coulomb, and exchange matrices obtained from the Hartree-Fock calculation. The general idea is that $s^{k, S}_{ll}(\ibo_{i}^{\mol{A}})$ includes information about momentum distribution of $\ibo_{i}^{\mol{A}}$, $s^{k, M}_{l_{1}l_{2}}(\ibo_{i}^{\mol{A}})$ terms for $M=F,J,K$ include information about averaged out interactions experienced by atom $k$, and $o^{k, M}_{l_{1}l_{2}}(\ibo_{i}^{\mol{A}})$ terms include information about the most chemically relevant interactions between atom $k$ and other atoms.

To summarise, we represent each localized orbital $\ibo_{i}^{\mol{A}}$ in a Slater determinant $\mol{A}$ in terms of pairs of $\tilde{\weight}_{k}(\ibo_{i}^{\mol{A}})$ and $\vecrep^{k}(\ibo_{i}^{\mol{A}})$. We will now explain how these are used to define kernel functions that can be used in kernel ridge regression.

\subsection{Kernel function for electronic structure and changes of electronic structure}
\label{subsec:orbital_based_kernel}

To define kernel expressions used in this work, we followed a philosophy largely similar to the one of  Refs.~\onlinecite{Bartok_Csanyi:2013,Ferre_Stoltz:2015,Glielmo_DeVita:2017}. We started by considering a sum of Gaussian functions centered on $\vecrep^{k}(\ibo_{i}^{\mol{A}})$ with coefficients $\tilde{\weight}_{k}(\ibo_{i}^{\mol{A}})$; following Refs.~\onlinecite{Bartok_Csanyi:2013,Ferre_Stoltz:2015} we define linear kernel element for two localized orbitals $\ibo_{i}^{\mol{A}}$ and $\ibo_{j}^{\mol{B}}$ as the overlap between two such functions
\begin{equation}
\begin{split}
    \Klin(\ibo_{i}^{\mol{A}}, \ibo_{j}^{\mol{B}})=&\sum_{k=1}^{\Natom^{\mol{A}}}\sum_{m=1}^{\Natom^{\mol{B}}}\frac{\tilde{\weight}_{k}(\ibo_{i}^{\mol{A}})\tilde{\weight}_{m}(\ibo_{j}^{\mol{B}})}{\prod_{p=1}^{\dimf}\sigma_{p}}\left(\frac{\pi}{2}\right)^{\dimf/2}\\
    &\times\exp\left\{-\sum_{p=1}^{\dimf}\frac{[\vecrep_{p}^{k}(\ibo_{i}^{\mol{A}})-\vecrep_{p}^{m}(\ibo_{j}^{\mol{B}})]^{2}}{4\sigma_{p}^{2}}\right\},\label{eq:Klin}
\end{split}
\end{equation}
where $\sigma_{p}$ ($p=1,\ldots,\dimf$) are hyperparameters.

Using $\Klin$ as a vector product, we can define the corresponding distance between two orbital representations. We divide all $\tilde{\weight}_{k}(\ibo_{i}^{\mol{A}})$ and $\tilde{\weight}_{m}(\ibo_{j}^{\mol{B}})$ by $\sqrt{\Klin(\ibo_{i}^{\mol{A}},\ibo_{i}^{\mol{A}})}$ and $\sqrt{\Klin(\ibo_{j}^{\mol{B}},\ibo_{j}^{\mol{B}})}$, then calculate the distance between two resulting representations in order to define the corresponding Gaussian kernel function\cite{Glielmo_DeVita:2017} which takes the form
\begin{equation}
\begin{split}
  \KGauss(\ibo_{i}^{\mol{A}}, \ibo_{j}^{\mol{B}})&=\exp\left\{-\frac{1}{\sigmaglobal^{2}}\vphantom{\left[1-\frac{\Klin(\ibo_{i}^{\mol{A}}, \ibo_{j}^{\mol{B}})}{\sqrt{\Klin(\ibo_{i}^{\mol{A}}, \ibo_{j}^{\mol{B}})\Klin(\ibo_{i}^{\mol{A}}, \ibo_{j}^{\mol{B}})}}\right]}\right.\\
 &\times\left.\left[1-\frac{\Klin(\ibo_{i}^{\mol{A}}, \ibo_{j}^{\mol{B}})}{\sqrt{\Klin(\ibo_{i}^{\mol{A}}, \ibo_{i}^{\mol{A}})\Klin(\ibo_{j}^{\mol{B}}, \ibo_{j}^{\mol{B}})}}\right]\right\},\label{eq:KGauss}
\end{split}
\end{equation}
where $\sigma_{\mathrm{global}}$ is a hyperparameter.

We define kernel function for two Slater determinants $\mol{A}$ and $\mol{B}$ as
\begin{equation}
    \Kmol(\mol{A},\mol{B})=\sum_{i=1}^{\Nocc^{\mol{A}}}\sum_{j=1}^{\Nocc^{\mol{B}}}\KGauss(\ibo_{i}^{\mol{A}},\ibo_{j}^{\mol{B}}).
    \label{eq:Kmol}
\end{equation}
The expression is similar to kernel functions used for modelling total potential energies as sums of atomic contributions\cite{Bartok_Csanyi:2010} and can be used for modelling extensive properties. In this work, we used it to estimate total potential energy corresponding to the ground state or a spin excited state of a molecule with $\mol{A}$ and $\mol{B}$ being Slater determinants obtained from a ground state or a spin excited state calculation. For intensive properties that were the focus of this work, namely HOMO and LUMO energies, ionization potentials, and electron affinities, we note that they can be qualitatively described as a change of energy between two Slater determinants. We define a kernel function corresponding to changes of Slater determinants from $\mol{A}_{1}$ to $\mol{A}_{2}$ and from $\mol{B}_{1}$ to $\mol{B}_{2}$, denoted as $\dpair{A}{1}{2}$ and $\dpair{B}{1}{2}$, as
\begin{equation}
\begin{split}
    \Kmol(\dpair{A}{1}{2}, \dpair{B}{1}{2})=&\Kmol(\mol{A}_{1}, \mol{B}_{1})+\Kmol(\mol{A}_{2},\mol{B}_{2})\\
    &-\Kmol(\mol{A}_{1},\mol{B}_{2})-\Kmol(\mol{A}_{2}, \mol{B}_{1}).\label{eq:Kpair}
\end{split}
\end{equation}
It is important to underline that while the representations of pairs $\dpair{A}{1}{2}$/$\dpair{B}{1}{2}$ include information about the entirety of Slater determinants $\mol{A}_{1}$/$\mol{B}_{1}$ and $\mol{A}_{2}$/$\mol{B}_{2}$, the kernel function~(\ref{eq:Kpair}) effectively acts on the corresponding changes of Slater determinants as localized orbitals not affected by the changes $\dpair{A}{1}{2}$/$\dpair{B}{1}{2}$ do not contribute to the expression. As a result, as long as the changes $\dpair{A}{1}{2}$ and $\dpair{B}{1}{2}$ qualitatively correctly reflect which part of a molecule is affected most by an electronic process, the latter will also contribute most to the kernel element.

Changes $\dpair{A}{1}{2}$ used in this work are illustrated in Figure~\ref{fig:Slater_determinant_change}. They were defined for different quantities of interest as:
\begin{itemize}
    \item \emph{HOMO energy}: change of HOMO occupation number from 2 to 1.
    \item \emph{LUMO energy}: change of LUMO occupation number from 0 to 1.
    \item \emph{First excitation}: change of HOMO and LUMO occupation numbers from 2 to 1 and from 0 to 1 without changing the total spin of the molecule.
    \item \emph{Ionization potential}: decrease the number of electrons by 1.
    \item \emph{Electron affinity}: increase the number of electrons by 1.
\end{itemize}

Now that we have defined kernel functions both for electronic structures and electronic processes, the last aspect of our method to be discussed is its coupling to $\Delta$-ML methods. For total potential energies, it always makes physical sense to apply FJK to corrections of {\it ab initio} calculations rather than the energies themselves; it is easily shown by considering a molecule that has been stripped of all electrons ($N_{\mathrm{occ}}^{\mol{A}}=0$). However, for energy differences corresponding to changes of electronic structure without a change in atomic positions there is no implicit reason for combining FJK with $\Delta$-ML, and for some numerical experiments presented in Sec.~\ref{sec:results_discussion} combining FJK with $\Delta$-ML increased prediction errors of the former.

As mentioned earlier, FJK has several conceptual similarities with MOB-ML; the latter method estimates correlation energy correction to the Hartree-Fock energy by evaluating contributions to this correction from pairs of localized orbitals using kernel ridge regression. The principal philosophical difference of FJK is that, while MOB-ML has been tailored to predicting correlation energy, we have avoided making any assumptions about the nature of error of the approximate estimate $p^{\mathrm{approx}}$; this was made to keep FJK as widely applicable as possible. A less important difference is that we represent localized orbitals rather than pairs of localized orbitals, making it easier to have information stored about a molecule scale linearly with its size; admittedly though, this scaling is made possible by neglecting parts of localized orbitals using $\tilde{\weight}_{k}(\ibo^{\mol{A}}_{i})$~(\ref{eq:cut_weights_def}). Interestingly, defining kernel elements corresponding to changes in electronic structure similarly to~(\ref{eq:Kpair}) might be possible in MOB-ML framework as well, and the resulting approach would benefit from the same cancellation of contributions from unaffected localized orbitals as FJK.

\section{Computational details}
\label{sec:global_comp_details}

 Our main goal was to test the  FJK approach with Hartree-Fock calculations which are as cheap as possible; therefore, unless stated otherwise, we used STO-3G basis,\cite{Hehre_Pople:1969,Hehre_Pople:1970} the resulting combination with FJK denoted FJK/HF/STO-3G. All Hartree-Fock calculations, as well as all orbital localizations, have been performed with the PySCF package;\cite{Sun:2015,Sun_Chan:2018,Sun_Chan:2020} for {\it ab initio} calculations where normal self-consistent field solver did not converge we used the second-order one described in Ref.~\onlinecite{Sun_Chan:2017}. We used $\rcut=0.05$ for cutting off negligible $\weight_{k}(\ibo_{i}^{\mol{A}})$.
 
 For each dataset considered in this work, we would first randomly choose a training set, then randomly and equally divide the remaining molecules between the validation and the test sets; we would first optimize hyperparameters using the training and validation sets, then use those hyperparameters to train models with the training set and calculate mean absolute errors (MAEs) over the test set, which are presented in this work. The learning curves were built by averaging MAEs obtained by training machine learning models on randomly chosen subsets of the training set of a given size (up to 8 such subsets were chosen for a given point). To avoid optimization over all $\sigma_{p}$ values we made them proportional to a rescaling hyperparameter that was optimized instead. We also found it challenging to locate value of $\lambda$ optimal for FJK, resorting to a complicated procedure that combined scanning with bisection for $\lambda$ optimization. The resulting hyperparameter optimization protocol went as follows:
\begin{itemize}
    \item Evaluate $\sigma_{p}^{0}$ ($p=1,\ldots, \dimf$) as
    \begin{equation}
    \begin{split}
        [\sigma_{p}^{0}]^{2}=&\sum_{t=1}^{\Ntrain}\sum_{i=1}^{\Nocc^{\mol{A}_{t}}}\sum_{k=1}^{\Natom^{\mol{A}_{t}}}\tilde{\weight}_{k}(\ibo_{i}^{\mol{A}_{t}})\left[\vecrep_{p}^{k}(\ibo_{i}^{\mol{A}_{t}})\right]^{2}\\
        &-\left[\sum_{t=1}^{\Ntrain}\sum_{i=1}^{\Nocc^{\mol{A}_{t}}}\sum_{k=1}^{\Natom^{\mol{A}_{t}}}\tilde{\weight}_{k}(\ibo_{i}^{\mol{A}_{t}})\vecrep_{p}^{k}(\ibo_{i}^{\mol{A}_{t}})\right]^{2},
    \end{split}
    \end{equation}
    where $\mol{A}_{t}$ is the ground state Slater determinant of the molecule in the training set with index $t$ ($t=1,\ldots,\Ntrain$).
    \item Introduce a rescaling hyperparameter $\sigmaresc$ and use $\sigma_{p}=\sigmaresc\cdot\sigma_{p}^{0}$.
    \item We calculate the MAE a model trained on the training set yields when applied to the validation set as a function of three hyperparameters: $\sigmaglobal$, $\sigmaresc$, and $\lambda$. To that end, we calculate kernel matrices for $\sigmaglobal$ and $\sigmaresc$ changing through $2^{-5},2^{-4},\ldots,2^{4},2^{5}$. For each of those precalculated matrices we optimized $\lambda$ by first scanning through values $10^{-9},10^{-8},\ldots,10^{-2},10^{-1}$ and then checking whether the MAE could be improved by choosing $\lambda$ from bisection-based minimization of root mean square error over the validation set.
\end{itemize}
 For QM7b and QM9 datasets, we additionally carried out kernel ridge regression calculations using Coulomb Matrix\cite{Rupp_Lilienfeld:2012,Ramakrishnan_Lilienfeld:2015_CM} and SLATM\cite{Huang_Lilienfeld:2020} representations with and without $\Delta$-ML; we optimized hyperparameters using the same protocol as FJK, except we scanned over Laplacian and Gaussian kernels\cite{Muller_Tsuda:2020} with $\sigma$ values varying through $1, 2, \ldots, 4096, 8192$.
 
As discussed in Subsec.~\ref{subsec:qm9_results_discussion}, when training FJK on 110k QM9 molecules we had to discard molecules that yielded representations of Slater determinants or pairs of Slater determinants that were close to being redundant. For Slater determinants, the algorithm used square distance defined as
\begin{equation}
    \Dsq(\mol{A},\mol{B})=\Kmol(\mol{A}, \mol{A})+\Kmol(\mol{B},\mol{B}) -2\Kmol(\mol{A},\mol{B}).
\end{equation}
We would find the pair $\mol{A}_{\mathrm{min}}$ and $\mol{B}_{\mathrm{min}}$ that has the minimal square distance among all pairs in the dataset, then randomly eliminate one of those entries from the training set; the procedure would be repeated until a predetermined number of entries has been discarded. The procedure worked analogously for pairs of Slater determinants.

  Some dataset-specific details of the calculations are left for Section~\ref{sec:datasets}.

\section{Datasets}
\label{sec:datasets}

\subsection{QM9}

QM9 dataset\cite{Ruddigkeit_Reymond:2012,Ramakrishnan_Lilienfeld:2014} contains 134k organic molecules with up to 9 heavy atoms (C, N, O, and F), with several properties calculated using B3LYP\cite{Stephens_Frisch:1994} density functional and 6-31G(2df,p)\cite{Hehre_Pople:1972,Krishnan_Pople:1980,Frisch_Pople:1984} basis set. The training set size was 110k molecules, during hyperparameter scan we trained tested models not over the entire training set, but over a random subset of the training set that had the same size as the validation set (11,943 molecules).

\subsection{QM7b}

QM7b\cite{Blum_Reymond:2009,Montavon_Lilienfeld:2013} dataset contains 7,211 organic molecules with up to 7 heavy atoms (C, N, O, S, and Cl), with a wide array of properties, some calculated with different levels of theory. The training set size was 5,000 molecules.

\subsection{QM7b-T and GDB-13-T dataset}

QM7b-T\cite{Welborn_Miller:2018,Cheng_Miller:2019a} contains thermalized configurations of molecules from QM7b, while GDB-13-T\cite{Welborn_Miller:2018,Cheng_Miller:2019a} was composed by taking 6 thermalized configurations from each of randomly chosen 1,000 molecules with 13 heavy atoms (C, N, O, S, and Cl, just as in QM7b and QM7b-T) from the GDB-13 dataset.\cite{Blum_Reymond:2009} Both datasets include total potential energies calculated at different levels of theory; since they have the same elemental composition but differ in sizes of included molecular conformers, they are often used to benchmark both a model's performance in estimating total potential energy values of smaller molecules (QM7b-T) and to check the model's transferability between smaller and larger molecules (GDB-13-T). The training set size for QM7b-T was 4,000 molecules.

\subsection{LIBE}

LIBE\cite{Spotte-Smith_Persson:2021} includes 17,190 species of different charges (-1, 0, and 1) and spin multiplicities (1, 2, and 3) containing up to 16 heavy atoms (Li, C, N, O, F, P, and S), making it significantly more diverse than the other datasets considered in this work. The data is calculated with $\omega$B97X-V\cite{Mardirossian_Head-Gordon:2014} density functional and def2-TZVPPD\cite{Weigend_Ahlrichs:2005,Rappoport_Furche:2010} basis set, with SMD\cite{Marenich_Truhlar:2009} used to simulate interaction with a Li-ion electrolyte solvent.

To divide LIBE into training, validation, and test datasets, we first sorted LIBE's species by their molecular graphs, then sorted those graphs according to which combinations of charge and spin they are represented with in LIBE. The latter groups of molecular graphs were then randomly divided between training, validation, or test sets with approximate ratios of 60\%, 20\%, and 20\%.

\section{Results and discussion}
\label{sec:results_discussion}

\subsection{QM9}
\label{subsec:qm9_results_discussion}

We applied FJK to total potential energies, HOMO and LUMO energies, and the HOMO-LUMO gap. We originally considered modelling the HOMO-LUMO gap using the Slater pair representation of first excitation [see Figure~\ref{fig:Slater_determinant_change} (c)], but we found this approach to yield larger errors than separate estimation of HOMO and LUMO energies with the gap calculated as their difference. We encountered difficulties producing results for 110k training set sizes since our hyperparameter optimization procedure led to situations where inversion $(\mat{K}+\lambda\mat{I})$ from Eq.~(\ref{eq:alpha}) was numerically unstable; this is probably a case of overfitting $\lambda$ for training sets significantly smaller than 110k molecules. The problem, however, did not affect performance of different kernel ridge regression methods in the small training data set limit which is the relevant limit for the motivation of this work. 

\begin{figure}
\includegraphics[width=0.475\textwidth]{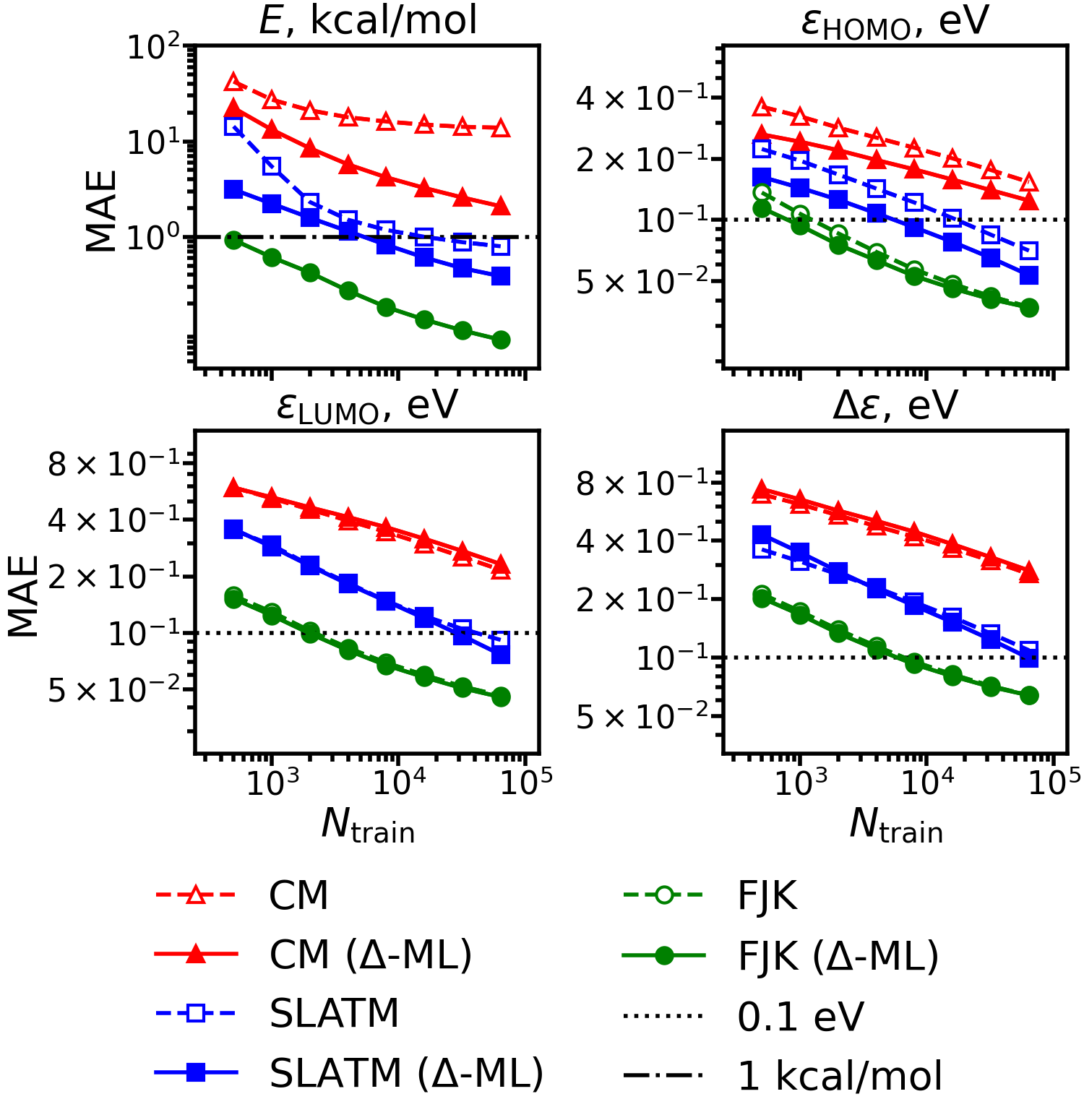}
\caption{\label{fig:qm9_learning_curves}Mean absolute errors (MAEs) for total potential energies ($E$), HOMO and LUMO energies ($\epsilon_{\mathrm{HOMO}}$ and $\epsilon_{\mathrm{LUMO}}$), and HOMO-LUMO gaps ($\Delta\epsilon$) of QM9 molecules obtained with Coulomb Matrix (CM), SLATM, and FJK/HF/STO-3G representations as a function of training set size, with and without $\Delta$-ML.}
\end{figure}

The learning curves for training set sizes up to 64k molecules are presented in Figure~\ref{fig:qm9_learning_curves} with results for Coulomb Matrix and SLATM added for comparison. FJK representation performs reasonably well for smaller training set sizes, reaching a mean absolute prediction error of 0.1~eV at less then 1,000, 2,000, and 8,000 training molecules respectively for HOMO energies, LUMO energies, and the HOMO-LUMO gap; while for the total potential energy the coveted threshold of chemical accuracy (MAE = 1~kcal/mol) is reached for less than 500 training molecules. Interestingly, comparing results with and without $\Delta$-ML for all three representations indicates that HF/STO-3G calculations would be of little use for predicting LUMO energy and HOMO-LUMO gap with a conventional $\Delta$-ML protocol. However, the electronic structure details coming from the same base-line calculations still seem to reflect the molecular systems' physics well enough to yield a well-performing FJK representation. A possible explanation for this is found in Ref.~\onlinecite{Mazouin_Lilienfeld:2021}, which argued that a major problem with using conventional representations for ML of HOMO and LUMO energies is lack of smoothness of the quantity with respect to the molecule's representation when molecules with radically different functional group compositions are considered. The authors bypassed the problem by introducing a classification of molecules according to their chemical groups into the ML protocol, significantly improving data efficiency. In this work, the smoothness problem is bypassed in the choice of kernel function~(\ref{eq:Kpair}), which, as discussed in Subsec.~\ref{subsec:orbital_based_kernel}, ensures that as long as the {\it ab initio} method used to generate FJK representation correctly identifies which functional group contributes most to HOMO and LUMO this group will yield the greatest contribution to the kernel function. Therefore, the accuracy of FJK/HF/STO-3G for LUMO energy and HOMO-LUMO gap predictions may indicate that HF/STO-3G's ability to qualitatively correctly locate a molecule's HOMO and LUMO affects FJK's performance more than its inability to accurately predict LUMO energy and HOMO-LUMO gap.

To obtain results for FJK representation at 110k molecules, we applied preliminary filtering of molecules with the procedure described in the end of Subsec.~\ref{sec:global_comp_details}; discarding 5,000 molecules in this way proved sufficient to negate the numerical instability issue outlined in the beginning of this subsection. The results are presented in Table~\ref{tab:QM9_results}; to put these numbers in perspective, we added MAEs yielded by DeepMoleNet,\cite{Liu_Ma:2021} SphereNet,\cite{Liu_Ji:2021} and PaiNN\cite{Schutt_Gastegger:2021} when trained on 110k training molecules. Those approaches use neural networks to predict quantities of interest and were chosen since they have demonstrated the best performance among the field for at least one quantity considered in this work. We also add MAEs yielded by OrbNet\cite{Qiao_Miller:2020} and UNiTE,\cite{Qiao_Miller:2021} two approaches that used neural networks on representations based on orbitals and operators obtained from GFN1-xTB\cite{Grimme_Shushkov:2017} calculations. We can see that in this setup, FJK proves to be very accurate for evaluating total potential energies, while in terms of accuracy in evaluating HOMO and LUMO energies, as well as HOMO-LUMO gaps, FJK only approaches the current state-of-the-art in the field.

\begin{table*}
\caption{Mean absolute errors (MAEs) yielded by FJK with and without $\Delta$-ML for QM9 dataset when trained on 110k molecules; we use the same names of quantities as in caption of Figure~\ref{fig:qm9_learning_curves}. MAEs for 110k training molecules yielded by some recent state-of-the-art approaches are added for comparison. 
\hfill}\label{tab:QM9_results}
\begin{ruledtabular}
\begin{tabular}{lccccccc}
\multirow{3}{*}{Property} & \multicolumn{7}{c}{MAE (meV)}\\
\cline{2-8}
 & \multirow{2}{*}{DeepMoleNet\cite{Liu_Ma:2021}}
 & \multirow{2}{*}{SphereNet\cite{Liu_Ji:2021}} & \multirow{2}{*}{PaiNN\cite{Schutt_Gastegger:2021}} & \multirow{2}{*}{OrbNet\cite{Qiao_Miller:2020}$^{,}$\footnote[1]{GFN1-xTB\cite{Grimme_Shushkov:2017} calculations used for reference\label{footnote:GFN1-xTB_ref}}}& \multirow{2}{*}{UNiTE\cite{Qiao_Miller:2021}$^{,}$\footref{footnote:GFN1-xTB_ref}} & \multicolumn{2}{c}{FJK/HF/STO-3G}\\
\cline{7-8}
& & & & & & no $\Delta$-ML& with $\Delta$-ML\\
\hline
$E$ & \phantom{0}6.1& \phantom{0}6.3 & \phantom{0}5.9 & 3.9 & \phantom{0}3.5 & \_ & \phantom{0}$3.2$ \\
$\epsilon_{\mathrm{HOMO}}$ & 21.9 & 23.6 & 27.6 & \_ & \phantom{0}9.9 & 33.9 & 34.5 \\
$\epsilon_{\mathrm{LUMO}}$ & 18.5 & 18.9 & 20.4 & \_ & 12.7 & 43.3 & 42.4 \\
$\Delta\epsilon$ & 32.1 & 32.3 & 45.7 & \_ & 17.3 & 59.0 & 59.5\\
\end{tabular}
\end{ruledtabular}

\end{table*}

\subsection{QM7b}
\label{subsec:QM7b_results}

\begin{figure*}
[tbp]%
\centering\includegraphics[width=1.0\textwidth]{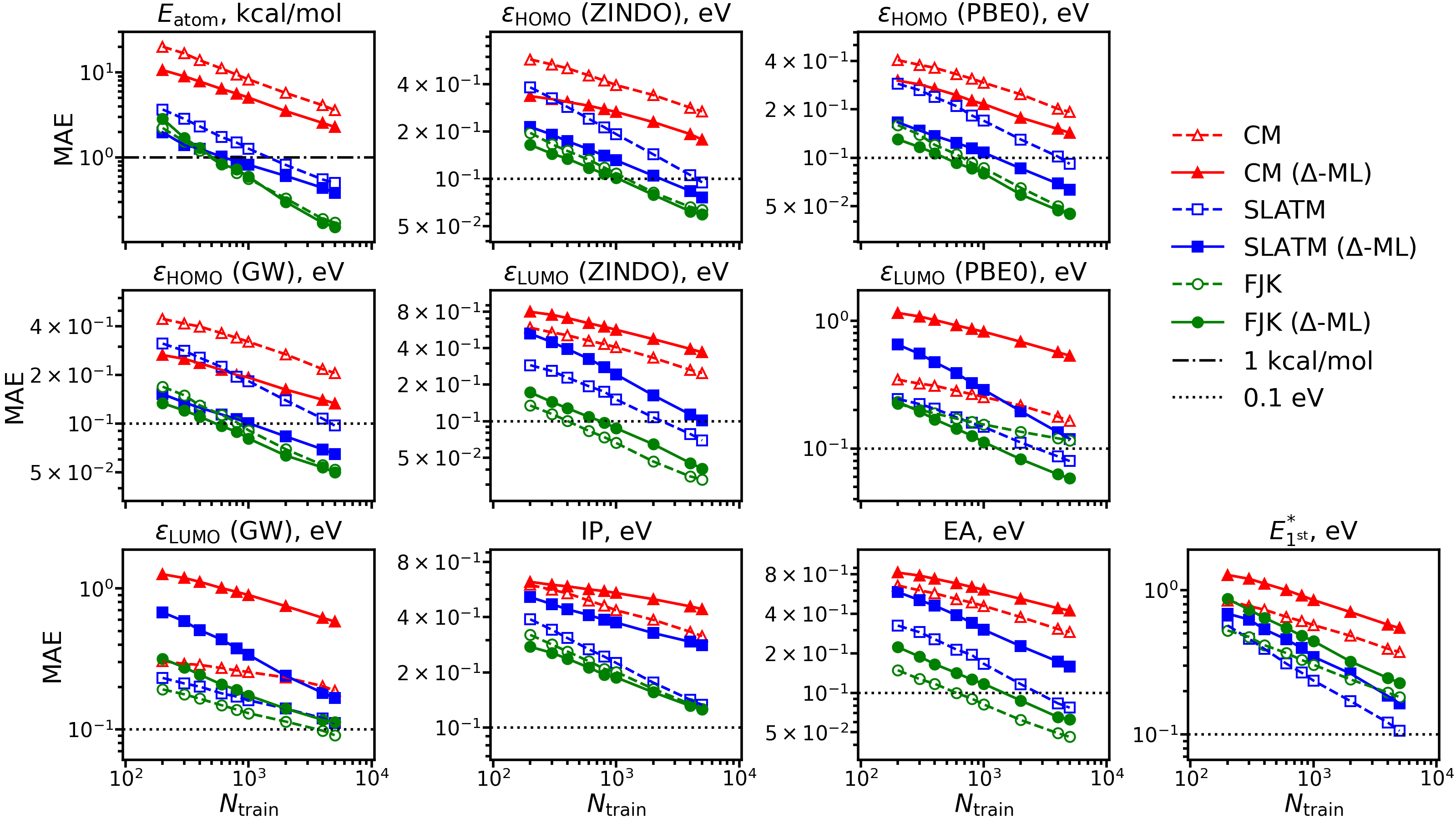}
\caption{\label{fig:qm7b_learning_curves}Mean absolute errors (MAEs) for atomization energies ($E_{\mathrm{atom}}$), HOMO and LUMO energies ($\epsilon_{\mathrm{HOMO}}$ and $\epsilon_{\mathrm{LUMO}}$) calculated at different levels of theory, ionization potentials (IP), electron affinities (EA), and first excitation energies ($E_{1^{\mathrm{st}}}^{*}$) of QM7b dataset as a function of the size of the training set. The results are obtained with Coulomb Matrix (CM), SLATM, and FJK/HF/STO-3G representations with and without $\Delta$-ML.}
\end{figure*}

For QM7b, we tested FJK's performance on machine learning atomization energies, HOMO and LUMO energies, ionization potentials, electron affinities, and first excitation energies; the resulting learning curves are plotted on Figure~\ref{fig:qm7b_learning_curves}, with MAEs for the largest training set size of 5,000 molecules presented in Table~\ref{tab:QM7b_results}. We can see that in this setup FJK tends to perform better than conventional representations proposed previously, though FJK's efficiency relative to other methods seems to depend greatly on how well a given quantity is represented by the Slater determinant pair used for the mapping. The quantity where FJK performs worse than many conventional representations is first excitation energy (where for $\Delta$-ML we used the HOMO-LUMO gap as the estimate), which is unsurprising given that transporting an electron from HOMO to LUMO [see Figure~\ref{fig:Slater_determinant_change}~(c)] is a very crude approximation to the excitation process. Interestingly, for some quantities using HF/STO-3G as a baseline in a $\Delta$-ML protocol tends to decrease accuracy of a ML model, but FJK/HF/STO-3G still performs significantly better than conventional representations; this is similar to the situation with LUMO and HOMO-LUMO gap predictions for QM9 discussed in Subsec.~\ref{subsec:qm9_results_discussion}.

\begin{table*}
\caption{Mean absolute errors (MAEs) yielded by FJK/HF/STO-3G with and without $\Delta$-ML when trained on 5,000 molecules. For comparison we present results previously reported for Coulomb Matrix (CM), SOAP, and Encoded Bonds\cite{Collins_Yaron:2018} (EB); Ref.~\onlinecite{Huang_Lilienfeld:2016} tested a number of representations and we present the best result among those. The convention for naming quantities is the same as in caption of Figure~\ref{fig:qm7b_learning_curves}.\hfill}\label{tab:QM7b_results}
\begin{ruledtabular}
\begin{tabular}{lcccccc}
\multirow{3}{*}{Property} & \multicolumn{6}{c}{MAE (eV)}\\
\cline{2-7}
 & \multirow{2}{*}{CM\cite{Montavon_Lilienfeld:2013}} & 
 \multirow{2}{*}{SOAP\cite{De_Ceriotti:2016}}&
 \multirow{2}{*}{Ref.~\onlinecite{Huang_Lilienfeld:2016}}&
 \multirow{2}{*}{EB\cite{Collins_Yaron:2018}} & \multicolumn{2}{c}{FJK/HF/STO-3G} \\
\cline{6-7}
& & &  &  & no $\Delta$-ML & with $\Delta$-ML \\
\hline
$E_{\mathrm{atom}}$
& $0.16$ & $0.04$ & $0.05$ & $ 0.04$ & $0.0075$ & $0.0066$ \\
$\epsilon_{\mathrm{HOMO}}$ (GW)
&   $0.16$ & $0.12 $ & $0.10$ & $ 0.13$ & $0.052$\phantom{0} & $0.050$\phantom{0}\\
$\epsilon_{\mathrm{HOMO}}$ (PBE0)
& $0.15$ & $0.11$ & \_ & $ 0.12$ & $0.045$\phantom{0} & $0.045$\phantom{0}\\
$\epsilon_{\mathrm{HOMO}}$ (ZINDO)
& $0.15$ & $0.13$ & \_ & $ 0.13$ & $0.064$\phantom{0} & $0.059$\phantom{0}\\
$\epsilon_{\mathrm{LUMO}}$ (GW)
& $0.13$ & $0.12$ & $0.11$ & $ 0.13$ & $0.091$\phantom{0}& $0.11$\phantom{00}\\
$\epsilon_{\mathrm{LUMO}}$ (PBE0) &
$0.12$ & $0.08$ & \_ & $ 0.09$ & $0.12$\phantom{00} & $0.058$\phantom{0}\\
$\epsilon_{\mathrm{LUMO}}$ (ZINDO)&
$0.11$ & $0.10 $ & \_ & $ 0.10$ & $0.033$\phantom{0} & $0.041$\phantom{0}\\
IP &
$0.17$ & $0.19$ & $0.15$ & $ 0.16$ & $0.13$\phantom{00} & $0.13$\phantom{00}\\
EA &
$0.11$ & $0.13$ & $ 0.07 $ & $ 0.09$ & $0.046$\phantom{0} & $0.063$\phantom{0}\\
$E^{*}_{\mathrm{1^{st}}}$
& $0.13$ & $0.18$ & $0.13$ & $0.23$ & $0.18$\phantom{00} & $0.23$\phantom{00}
\end{tabular}
\end{ruledtabular}

\end{table*}

\subsection{QM7b-T and GDB-13-T datasets}

\begin{figure}
[tbp]%
\centering\includegraphics[width=0.475\textwidth]{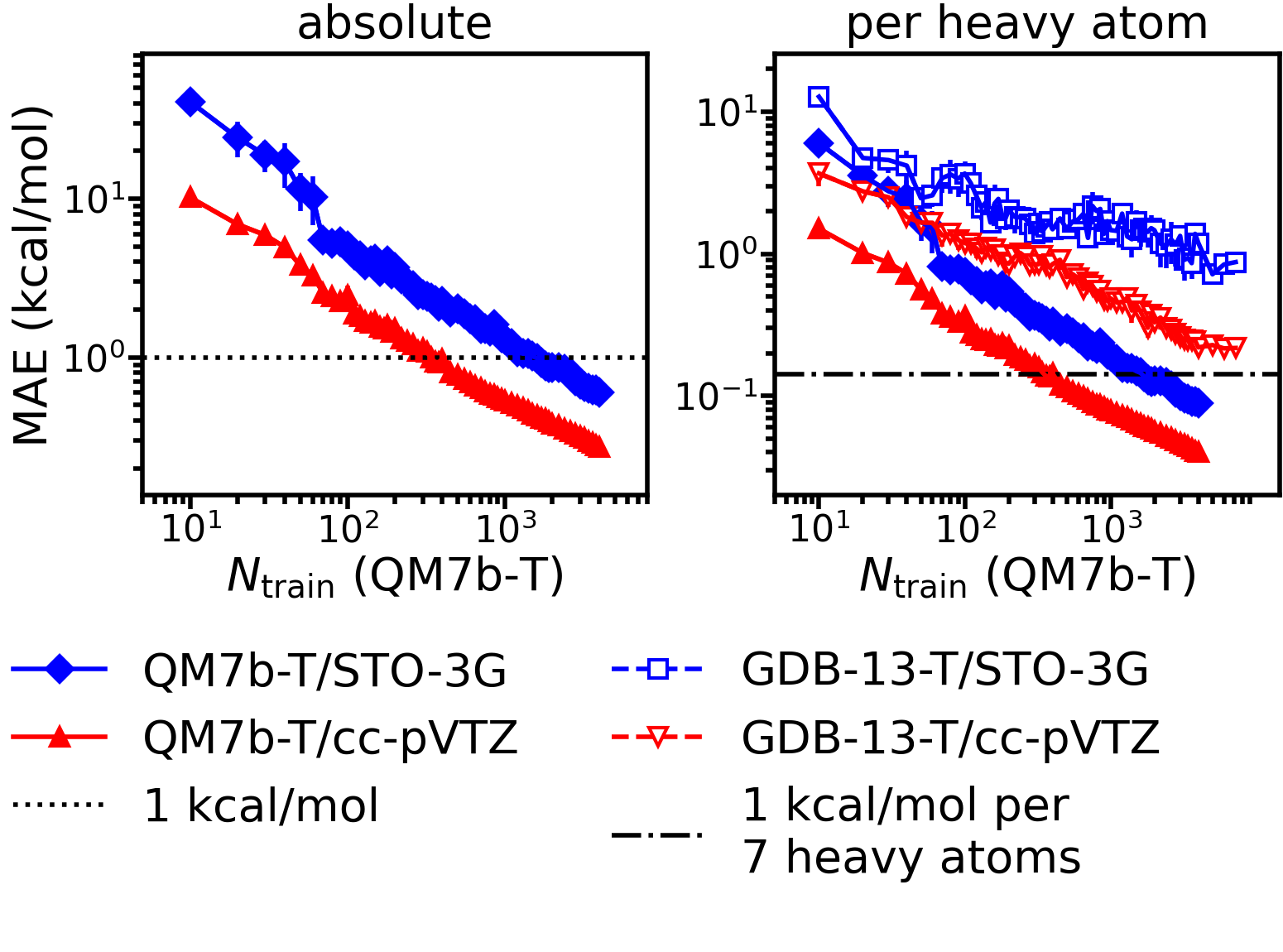}
\caption{\label{fig:qm7bt_learning_curves} Mean absolute errors (MAEs) yielded by FJK models trained on QM7b-T configurations for MP2/cc-pVTZ potential energies of QM7b-T and GDB-13-T configurations with HF/STO-3G or HF/cc-pVTZ calculations used as reference.}
\end{figure}

We tested performance of FJK/HF/STO-3G and FJK with cc-pVTZ\cite{Dunning:1989,Woon_Dunning:1993} basis Hartree-Fock calculations as reference (FJK/HF/cc-pVTZ) for predicting MP2/cc-pVTZ level QM7b-T's total potential energies. Following previous work, we also checked FJK's transferability by applying models trained on QM7b-T to MP2/cc-pVTZ total potential energies of GDB-13-T. The resulting learning curves are presented in Figure~\ref{fig:qm7bt_learning_curves}.

For predicting QM7b-T total potential energies, FJK/HF/STO-3G and FJK/HF/cc-pVTZ pass the chemical accuracy threshold of 1~kcal/mol at 1600 and 340 training molecules; in Table~\ref{tab:QM7bT_chem_acc} these numbers are compared to the ones reported for MOB-ML,  FCHL,\cite{Faber_Lilienfeld:2018} FCHL19,\cite{Christensen_Lilienfeld:2020} and the combination of FCHL with $\Delta$-ML using HF/cc-pVTZ as reference. We can see that in this setup FJK/HF/STO-3G performs worse than both FCHL and FCHL19, which are in turn outperformed by FJK/HF/cc-pVTZ, though at the expense of requiring an expensive HF/cc-pVTZ calculation. Among other $\Delta$-ML protocols with HF/cc-pVTZ as reference, FJK/HF/cc-pVTZ outperforms combination of FCHL with conventional $\Delta$-ML, but is not as accurate as MOB-ML.

\begin{table}
\caption{Training set sizes required by different methods to reach 1~kcal/mol mean absolute error of predicting total potential energy [CCSD(T)/cc-pVTZ for DeePHF, MP2/cc-pVTZ for other methods] of QM7b-T configurations when trained on other QM7b-T configurations.\hfill}\label{tab:QM7bT_chem_acc}
\begin{ruledtabular}
\begin{tabular}{lc}
Method & $N_{\mathrm{train}}$ (1~kcal/mol)\\
\hline
FJK/HF/STO-3G & \alignedmultrow{5}{4.2}{1600, \\
360, \\
180,\cite{Cheng_Miller:2019a} \\
700,\cite{Cheng_Miller:2019a} \\
< 300,\cite{Chen_E:2020} \\
400-800,\cite{Christensen_Lilienfeld:2020}\\
400-800,\cite{Christensen_Lilienfeld:2020}} \\
FJK/HF/cc-pVTZ & \\
MOB-ML\footnote[1]{HF/cc-pVTZ calculations used for reference\label{footnote:hf_ccpvtz_ref}} &\\
FCHL/$\Delta$-ML\footref{footnote:hf_ccpvtz_ref} & \\
DeePHF\footref{footnote:hf_ccpvtz_ref} & \\
FCHL \\
FCHL19 \\
\end{tabular}
\end{ruledtabular}

\end{table}

To estimate transferability of FJK models from smaller to larger molecules, we compare MAEs normalized by the number of heavy atoms obtained for QM7b-T and GDB-13-T datasets when using FJK models trained on QM7b-T dataset. While for MOB-ML the resulting normalized MAEs have been shown to be rather similar between the two datasets,\cite{Cheng_Miller:2019a} for FJK they exhibit a drastic increase when going from QM7b-T to GDB-13-T. For a training set of 5,000 QM7b-T molecules, FJK/HF/STO-3G and FJK/HF/cc-pVTZ yield mean absolute errors of 9.33~and 2.96~kcal/mol; for comparison, in these conditions FCHL/$\Delta$-ML (with HF/cc-pVTZ as reference) yielded\cite{Cheng_Miller:2019a} mean absolute error of 2.43~kcal/mol.

\subsection{LIBE dataset}

We tested FJK's performance in several regimes. First, we predicted total potential energies as a general function of geometry, spin, and charge. Second, we singled out species with charge $0$ and spin multiplicity $1$, and compared how accurately their total potential energies can be predicted with models trained on only other such species, or on all types of species present in LIBE; that allowed us to see whether there is any synergy between those different classes of data. Lastly, we separately checked how our models performed for estimating energy differences associated with adiabatic ionization potentials, electron affinities, and $T_{1}$ excitation energies ({\it i.e.} differences of total potential energy of species corresponding to addition/removal of one electron with both oxidized and reduced species being in their lowest spin states, and to change of spin multiplicity without change of charge; the energy differences are between the corresponding potential minima).

\begin{figure}
[tbp]%
\includegraphics[width=0.475\textwidth]{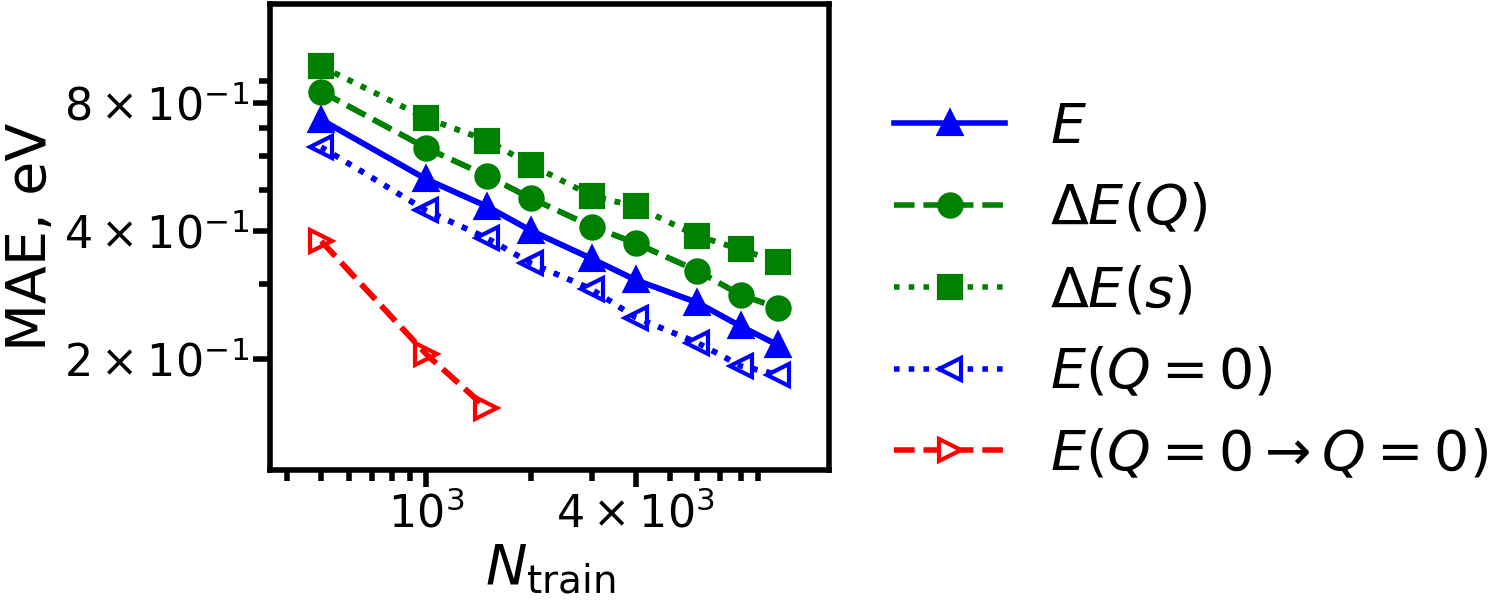}
\caption{\label{fig:libe_learning_curves}Mean absolute errors (MAEs) yielded by FJK/HF/STO-3G when used on LIBE for predicting total potential energies ($E$), as well as changes of total potential energies associated with change of charge [$\Delta E(Q)$] and spin [$\Delta E(s)$]; we also show mean absolute errors for predicting total potential energies of species of charge zero and spin multiplicity one when only such species are used for training [$E(Q=0\rightarrow Q=0)$] or all available types of species are used for training [$E(Q=0)$].}
\end{figure}

The resulting mean absolute errors are plotted as a function of training set size in Fig.~\ref{fig:libe_learning_curves}, with the values obtained for largest training set sizes gathered in Table~\ref{tab:LIBE_results}. We can see that despite the complexity of the dataset and the reference HF/STO-3G calculations not including solvent effects, FJK/HF/STO-3G exhibits adequate learning rates for total potential energies and for energy differences considered in this work. Mean absolute errors for energies of neutral species of spin multiplicity $1$ are $\sim$20\% smaller if only other such species are used for training, meaning that FJK is currently not transferable enough to capitalize on synergies between different types of data in LIBE. However, the closeness of the two mean absolute error values also means that FJK is not significantly inhibited by the diversity of LIBE dataset either.

\begin{table}
\caption{Mean absolute errors (MAEs) yielded by FJK/HF/STO-3G for predicting LIBE's total potential energies and changes in total potential energy when the models are trained with the largest training set [1,487 species for $E(Q=0\rightarrow Q=0)$, 10,273 species for other quantities]. Same convention is used for quantity names as in the caption of Figure~\ref{fig:libe_learning_curves}.\hfill}\label{tab:LIBE_results}
\begin{ruledtabular}
\begin{tabular}{lc}
Property & MAE (eV)\\
\hline
$E$ & 0.216 \\
$\Delta E(Q)$ & 0.264 \\
$\Delta E(s)$ & 0.339 \\
$E(Q=0)$ & 0.184 \\
$E(Q=0\rightarrow Q=0)$ & 0.154
\end{tabular}
\end{ruledtabular}

\end{table}

\section{Computational timing and data efficiency}
\label{sec:computational_timing}

It should be pointed out that for a given training set size, FJK is significantly more computationally expensive to use than Coulomb Matrix or SLATM representations, due to both the need to run a Hartree-Fock calculation and the $ \mathcal{O}(\Nocc^{\mol{A}}\Nocc^{\mol{B}}) $ exponent evaluations needed to calculate one kernel element both for single Slater determinant~(\ref{eq:Kmol}) and a pair of Slater determinants~(\ref{eq:Kpair}). To compare computational time associated with using the three representations, we randomly chose two non-overlapping sets of 10k QM9 molecules, and used the time it took to generate the corresponding representations and to calculate a 10k$\times$10k kernel matrix for the two sets to estimate the time needed to evaluate properties for 10k QM9 molecules if the training set consists of 10k or 110k QM9 molecules. For FJK, we separately considered computational cost of evaluating the total potential energy and HOMO and LUMO energies since representation of a molecule with $N_{e}$ electrons includes distinct representations of $N_{e}/2$, $3N_{e}/2-1$, and $3N_{e}/2+1$ localized orbitals $\ibo_{i}^{\mol{A}}$, resulting in different costs of generating orbitals' representations and evaluating the corresponding kernel elements. In this setup, to generate a pair of Slater determinants necessary for evaluating HOMO and LUMO energies using FJK we performed HF/STO-3G and UHF/STO-3G calculations without reusing results from the former in the latter; this corresponds to the upper bound of how much computational time would be required to generate a Slater determinant pair corresponding to a given quantity. We also estimated the cost of running 10k calculations at B3LYP/6-31G(2df,p) level of theory ({\it i.e.}, the level of theory of QM9 data) by running pySCF calculations for 200 randomly chosen QM9 molecules. For Coulomb Matrix and SLATM, the computational timing tests were run with Gaussian kernel. All calculations were performed on an AMD Epyc 7,402 processor (24 cores, 512GB of RAM) using 8 parallel threads. For Coulomb Matrix and SLATM, we used Gaussian kernel.\cite{Muller_Tsuda:2020}

The results are presented in Table~\ref{tab:comp_timings}. It illustrates that if the reference data is abundant enough and the goal is to create a model that can predict a molecule's electronic properties with as little computational expense as possible, then conventional geometric representations of a molecule may be more suitable for this task. However, for smaller training sets the computational cost of using FJK is comparable to the computational cost of a combination of $\Delta$-ML with a conventional representation; this, combined with results presented in Subsec.~\ref{sec:results_discussion}, indicates that FJK would be the preferred method if the goal is to build an accurate model from scarce reference data.

\begin{table*}
\caption{Computational cost of using a model to make predictions for 10k QM9 molecules using models trained on 10k or 110k QM9 molecules, with or without using HF/STO-3G for $\Delta$-ML.  The cost of running 10k DFT calculations at QM9's level of theory [{\it i.e.} B3LYP/6-31G(2df,p)] was estimated as 332.4h.\hfill}\label{tab:comp_timings}
\begin{ruledtabular}
\begin{tabular}{ccccccc}
%\multirow{2}{*}{Training set size}
Training set size
& \multirow{2}{*}{used $\Delta$-ML}
 & \multirow{2}{*}{Coulomb Matrix} & \multirow{2}{*}{SLATM} & \multicolumn{3}{c}{FJK/HF/STO-3G}\\
\cline{5-7}
(QM9 molecules) & & & & $E$ & $\epsilon_{\mathrm{HOMO}}$ & $\epsilon_{\mathrm{LUMO}}$\\
\hline
\multirow{2}{*}{10k} & no & \phantom{0,00}8.3 s & \phantom{0,}242 s &
\_ & \_ & \_\\
%\cline{3-4}
 & yes & 1,886 s\phantom{.0} & 2,119 s & 4,846 s\phantom{.00} & \phantom{0}7.94 h & \phantom{0}8.12 h\\
 \hline
\multirow{2}{*}{110k} & no & \phantom{0,0}91 s\phantom{.0} & 2,311 s &
\_ & \_ & \_\\
%\cline{3-4}
& yes & 1,968 s\phantom{.0} & 4,188 s &
\phantom{0,00}6.60 h & 58.5 h\phantom{0} & 60.4 h\phantom{0}\\
\end{tabular}
\end{ruledtabular}

\end{table*}

To give a more quantitative estimate of how sparse reference data should be to make FJK preferable to conventional representations, we took the learning curves presented in Subsec.~\ref{subsec:qm9_results_discussion} for QM9 dataset and estimated from interpolation how many data points are required to train a model with MAE of 0.1~eV for HOMO and LUMO energies and 1~kcal/mol for total potential energy if SLATM or FJK representations are used (note that in our setup these target accuracies were not reached with Coulomb Matrix representation). The results are presented in Table~\ref{tab:dataset_comp_timings}, along with estimated computational times needed for models trained on these numbers of datapoints to make 10k predictions for QM9 molecules. According to these estimates, FJK/HF/STO-3G models would require between 21 and 37 smaller training set size to reach target accuracy, but would also take between 8 and 27 more computational time to produce the estimates in comparison to SLATM models.

\begin{table}
\caption{Estimated number of QM9 molecules necessary to achieve mean absolute error of 1~kcal/mol for total potential energy and 0.1~eV for HOMO and LUMO energies of QM9 molecules, as well as the estimated computational time necessary for a model trained on a dataset of this size to predict these quantities for 10k QM9 molecules.\hfill}\label{tab:dataset_comp_timings}
\begin{ruledtabular}
\begin{tabular}{ccccc}
%\multirow{2}{*}{Training set size}
\multirow{2}{*}{Quantity}
& \multicolumn{2}{c}{SLATM (no $\Delta$-ML)} &
\multicolumn{2}{c}{FJK/HF/STO-3G}\\
\cline{2-3}\cline{4-5}
& mol. num. & time & mol. num. & time\\
\hline
$E$ & 16,257 & 372 s & \phantom{0,}444\footnote{Estimated from extrapolation.} & 3,038 s\phantom{.00}\\
$\epsilon_{\mathrm{HOMO}}$ & 17,183 & 391 s & \phantom{0,}798\phantom{$^{\mathrm{a}}$} & \phantom{0,00}3.29 h\\
$\epsilon_{\mathrm{LUMO}}$ & 40,965 & 883 s & 1,953\phantom{$^{\mathrm{a}}$} & \phantom{0,00}3.92 h\\
\end{tabular}
\end{ruledtabular}

\end{table}

\section{Conclusions and future outlook}
\label{sec:conclusions_outlook}

We propose an orbital-based representation of molecular electronic structure, dubbed FJK, that can be used in conjunction with $\Delta$-ML approaches for estimating the total potential energy of a molecule, and extend it to mapping molecular processes associated with change of electronic structure onto pairs of Slater determinants obtained from computationally inexpensive Hartree-Fock calculations. Using QM7b and QM9 datasets as examples, we demonstrate how a good choice of such a mapping results in learning efficiency of the resulting models. Interestingly, this was even observed in situations where the \emph{ab initio} calculation used to generate the mapping was not accurate enough to be useful in conventional $\Delta$-ML protocols. We also demonstrated FJK's ability to be trained on very complex datasets containing species that differ in both charge and spin multiplicities by successfully applying the method to the LIBE data set. The FJK framework is general enough to be applicable to situations where magnitude of a complicated process can be estimated with a relatively simple orbital-based descriptor.\cite{Han_Cheng:2021,Hasebe_Tsuneda:2021} Further applications were beyond the scope of this work and are part of future work. 

At the same time, we observed that for a poor choice of a Slater determinant pair representing the change of electronic structure FJK may be less accurate than conventional representations, as exemplified by QM7b's first excitation energies. Also, when tested on QM7b-T and GDB-13-T datasets FJK demonstrated a smaller degree of transferability from smaller to larger molecules than MOB-ML and FCHL/$\Delta$-ML. In the future, we plan to investigate whether these issues can be avoided with another choice of localized orbitals or matrices $M$ used to define coupling functions in Eqs.~(\ref{eq:self_coupling}) and~(\ref{eq:other_coupling}). Lastly, we encountered numerical issues while pushing FJK training set sizes to larger numbers such as the 110k QM9 molecules. These issues  included stability as well as a rather prohibitive computational cost [as demonstrated in Sec.~\ref{sec:computational_timing}]. 
Ways of consistently mitigating these issues will also be investigated in future studies.

\section{Acknowledgments}

This project has received funding from the European Union’s Horizon 2020 research and innovation programme under grant agreement No 957189. The computational results presented have been achieved using the Vienna Scientific Cluster (VSC).

\section{Supplementary Information}

Raw data plotted in Figures presented in this work is available in Supplementary Information.

\section*{References}

\end{document}